\documentclass[12pt]{article}%
\usepackage{amsfonts}
\usepackage{amsmath}
\usepackage{amssymb}
\usepackage{graphicx}%
\providecommand{\U}[1]{\protect\rule{.1in}{.1in}}

\begin{document}

\title{The equations of Einstein and Cartan}
\author{D C Robinson\\Mathematics Department\\King's College London\\Strand, London WC2R 2LS\\United Kingdom\\email: david.c.robinson@kcl.ac.uk}
\maketitle

\begin{abstract}
A formulation of Einstein's gravitational field equations\ in four space-time
dimensions is presented using generalized differential forms and Cartan's
equations for metric geometries. \ Cartan's structure equations are extended
by using generalized metric connections. \ They are then employed to represent
Einstein's field equations and their solutions. \ When the energy-momentum
tensor is zero the generalized connections can be chosen to be flat and
different solutions of Einstein's equations can be related by generalizations
of the Poincar\'{e} group. \ An action \ for the vacuum field equations is
constructed by generalizing the Nieh-Yan three-form.

\end{abstract}

\section{Introduction}

This paper employs a generalization of \'{E}lie Cartan's formulation of the
exterior algebra and calculus of differential forms in a consideration of the
field equations of Albert Einstein's general relativity. \ Generalized
connections are used to extend Cartan's structure equations for metric
geometries. \ These latter equations are shown to be equivalent to Einstein's
gravitational field equations in four dimensional space-times\footnote{In this
paper vacuum equations means that the cosmological constant and
energy-momentum tensor are both zero.}. \ When such connections are flat they
can be used to describe the relationship between solutions, of Einstein's
equations. \ A first order action principle for Einstein's vacuum field
equations is constructed by replacing the ordinary connection in the Nieh-Yan
three-form by a flat generalized connection. \ This extends previous work that
used generalized forms to construct action principles.

The paper is organized in the following way. \ The next two review sections
are intended to make the paper reasonably self-contained. \ The second section
contains an extended review of the use of ordinary differential forms and
Cartan's structure equations in formulating Einstein's equations in four
dimensional space-times with Lorentzian signature metrics \cite{nak},
\cite{flanders}. \ This collection of results focuses on ones which are
generalized in subsequent sections and it also sets conventions that will be
used for both ordinary and generalized forms. \ The third section outlines the
properties of generalized differential forms that are needed in this paper.
\ In particular, type $N=2$ generalized forms, which are the ones used in
subsequent sections, are reviewed. \ More details about generalized
differential forms can be found in references \cite{sp1} -\cite{song2}. \ 

The fourth section employs the results from the previous three sections to
show how Einstein's gravitational field equations can be re-formulate by using
generalized connections and a generalization of Cartan's structure equations
for metric geometries. \ These connections extend ordinary connections and
covariant derivatives to generalized connections and the corresponding
covariant derivatives. \ Flat generalized connections in particular play a
central role in this paper. \ Einstein's equations, first in the vacuum case
and then in the case where the energy momentum tensor and/or the cosmological
constant are non-zero, are discussed. \ It is shown how to encode these
equations in generalizations of Cartan's structure equations. \ This is done
in the vacuum case by replacing the ordinary Levi-Civita connection one-form
in Cartan's first structure equations by a flat generalized connection. \ In
the non-vacuum case a generalized torsion two-form is added to represent the
cosmological constant and matter source terms. \ Following the procedure for
ordinary forms outlined in the second section these terms are then
incorporated into a modified generalized connection which has zero torsion.
\ These latter results are considered in greater detail when the cosmological
constant is non-zero but the energy-momentum tensor is zero. \ It is shown
that in this case, as in the vacuum case, the modified generalized connection
is flat. \ Those results in this section that apply to Einstein's vacuum field
equations have been previously described, using the formalism of two-component
spinors, \cite{pen1}, in a precursor to this paper, \cite{rob10}.

In the fifth section ordinary connections with values in the Lie algebra of
the Poincar\'{e} group are generalized by using results of the previous
section. \ Such connections can be regarded as generalizations of Cartan
connections (for a discussion of Cartan connections see \cite{wise}). \ It is
noted that when the energy-momentum tensor is zero these generalized
connections are flat. \ In the case where the cosmological constant is also
zero it is shown, by using a factorization, how to construct such connections
by combining certain complex conjugate flat generalized connections. \ Finally
in this section it is observed that solutions of Einstein's equations, with
either zero or non-zero cosmological constants can be related, in a unified
way, by generalizations of ordinary Poincar\'{e} transformations.

A Lagrangian for the Einstein vacuum field equations in four space-time
dimensions is constructed in the sixth section. \ Generalized forms have been
used in the construction of action principles for a variety of field theories,
\cite{tung1}, \cite{tung2}, \cite{rob6}, \cite{rob8}, \cite{rob9}. \ The
approach taken here focuses on the use of type $N=2$ forms and flat
generalized connections. \ It differs from previous results, where use was
made of a generalized Chern-Simons three-form, by employing a generalization
of the ordinary Nieh-Yan three-form \cite{N-Y}. The ordinary Levi-Civita spin
connection\ in that three-form is replaced by the flat generalized Levi-Civita
connection introduced in the fourth section. \ The action integral, in common
with other actions constructed using generalized forms, can include boundary terms.

In the last section the results of this paper are discussed and possible
applications of analogous approaches to other geometries and classical field
theories, such as four dimensional metric geometries with other signatures and
Yang-Mills fields are noted.

Conventions employed in previous papers are followed; in particular ordinary
forms are denoted by Greek letters, generalized forms and their derivatives by
bold Latin letters, lower case Latin superscripts and subscripts range and sum
from 1 to 4, Greek super- and sub-scripts range and sum from 1 to 5. \ In the
main local geometrical formulations are presented.

\section{Lorentzian four-metrics and Einstein's equations}

This section contains a review of the representations of Einstein's
gravitational field equations and Cartan's structure equations, for
four-metrics of signature $(1,3),$ which will be used in subsequent sections.

Lorentzian 4-metrics on a four dimensional space-time manifold $M$ can be
expressed in terms of an orthonormal co-frame of (ordinary) one forms
$\theta^{a}$ as
\begin{equation}
ds^{2}=\eta_{ab}\theta^{a}\otimes\theta^{b}%
\end{equation}
where\ $\eta_{ab}=diag(1,-1,-1,-1)$ $\footnote{Lower case Latin indices a,b,c,
etc are lowered and raised by $\eta_{ab}$ and its inverse in the usual way.}$.
\ These one-forms satisfy the first Cartan structure equations which can be
written as%
\begin{equation}
D\theta^{a}\equiv d\theta^{a}+\alpha_{b}^{a}\theta^{b}=d\theta^{a}+\text{
}^{-}\alpha_{b}^{a}\theta^{b}+\text{ }^{+}\alpha_{b}^{a}\theta^{b}=\Theta^{a}.
\end{equation}
Here $\Theta^a$ is the torsion two-form and $^-\alpha_b^a$ and $^+\alpha_b^a$
are respectively the anti self-dual and self dual parts of the metric
connection one-form $\alpha_b^a$ and are given by%
\begin{align}
^{+}\alpha_{b}^{a}  &  =\frac{1}{2}(\alpha_{b}^{a}-i\text{ }^{\ast}\alpha
_{b}^{a}),^{-}\alpha_{b}^{a}=\frac{1}{2}(\alpha_{b}^{a}+i\text{ }^{\ast}%
\alpha_{b}^{a}),\\
^{\ast}\alpha_{b}^{a}  &  =\frac{1}{2}\epsilon_{abcd}\alpha^{cd},\text{
}\alpha_{ab}=-\alpha_{ba},\nonumber
\end{align}
\ and the components of the totally skew symmetric Levi-Civita tensor are
\[
\epsilon_{abcd}\text{ where }\epsilon_{1234}=1.
\]

The second Cartan structure equations are%
\begin{equation}
\digamma_{b}^{a}=d\text{ }\alpha_{b}^{a}+\text{ }\alpha_{c}^{a}\text{ }%
\alpha_{b}^{c}=\frac{1}{2}\digamma_{bcd}^{a}\theta^{c}\theta^{d}%
\end{equation}
where $\digamma_{b}^{a}$ denotes the curvature two-form with components
$\digamma_{bcd}^{a}$. \ These curvature two-form can be expressed as the sum
of its anti self-dual and self-dual parts, $\digamma_{b}^{a}=$ $^{-}%
\digamma_{b}^{a}+$ $^{+}\digamma_{b}^{a}$ and%
\begin{align}
\digamma_{b}^{a}  &  =d\text{ }^{-}\alpha_{b}^{a}+\text{ }^{-}\alpha_{c}%
^{a}\text{ }^{-}\alpha_{b}^{c}=\frac{1}{2}\text{ }^{-}\digamma_{bcd}^{a}%
\theta^{c}\theta^{d},\\
^{+}\digamma_{b}^{a}  &  =d\text{ }^{+}\alpha_{b}^{a}+\text{ }^{+}\alpha
_{c}^{a}\text{ }^{+}\alpha_{b}^{c}=\frac{1}{2}\text{ }^{+}\digamma_{bcd}%
^{a}\theta^{c}\theta^{d},\nonumber
\end{align}
$\digamma_{abcd}$ is skew symmetric on the indices $a$ and $b$ and on the
indices $c$ and $d$ and similarly for the anti self-dual and self-dual components.

The first and second Bianchi identities are%

\begin{align}
\digamma_{b}^{a}\theta^{a}  &  =\ ^{-}\digamma_{b}^{a}\theta^{b}+\text{ }%
^{+}\digamma_{b}^{a}\theta^{b}=D\Theta^{a},\\
D\digamma_{b}^{a}  &  =0,\text{ }^{-}D^{-}\digamma_{b}^{a}=0,\text{ }^{+}%
D^{+}\digamma_{b}^{a}=0,\nonumber
\end{align}
where $D,$ $^{-}D$ and $^{+}D$ respectively denote the relevant exterior
covariant derivative with respect to the connection and anti self-dual or
self-dual part of the connection.

The second Cartan equations also imply that
\begin{equation}
D(^{\ast}\digamma_{b}^{a})=0,
\end{equation}
where the dual curvature two-form is $^{\ast}\digamma_{b}^{a}=\frac{1}%
{2}\epsilon_{bcd}^{a}\digamma^{cd}$.

The structure group is the proper, orthochronous Lorentz group $SO(1,3)$,
$\ L=$ $^{-}L\times$ $^{+}L$ $\sim SL(2,C)\times\overline{Sl(2,C)}$.

Its Lie algebra is $so(1,3),$ $l=($ $^{-}l\oplus$ $^{+}l)/Z_{2}\sim
(sl(2,C)\oplus$ $\overline{sl(2,C)})/Z_{2}$. \footnote{\ $^{+}L_{b}^{a}$ and
$^{-}L_{b}^{a}$ denote complex conjugate complex Lorentz transformations
isomorphic to spin transformations \cite{pen1} . \ In terms of two component
spinors $^{+}L_{b}^{a}\leftrightarrow\delta_{B}^{A}L_{B^{\prime}}^{A^{\prime}%
}$, $^{-}L_{b}^{a}\leftrightarrow\delta_{B^{\prime}}^{A^{\prime}}L_{B}^{A}%
$.where $L_{B}^{A}$ and its complex conjugate $L_{B^{\prime}}^{A^{\prime}}$
are $SL(2,C)$ elements satisfying $L_{C}^{A}$ $L_{D}^{B}\epsilon_{AB}%
=\epsilon_{CD}$ and its complex conjugate. \ The constant skew-symmetric
spinor\ $\epsilon_{AB}$ has $\epsilon_{01}=1.$ Furthermore $^{+}\alpha_{b}%
^{a}\leftrightarrow\delta_{B}^{A}\alpha_{B^{\prime}}^{A^{\prime}}$
,$^{-}\alpha_{b}^{a}\leftrightarrow\delta_{B^{\prime}}^{A^{\prime}}\alpha
_{B}^{A}$ , $^{+}\digamma_{b}^{a}\leftrightarrow\delta_{B}^{A}\digamma
_{B^{\prime}}^{A^{\prime}}$, $^{-}\digamma_{b}^{a}\leftrightarrow
\delta_{B^{\prime}}^{A^{\prime}}\digamma_{B}^{A}$. and the forms $\alpha
_{B}^{A}$, $\digamma_{B}^{A}$ and their complex conjugates take values in the
Lie algebra of $SL(2,C)$. \ Upper case Latin indices range and sum over 0 to
1.}. \ Hence%
\begin{align}
L_{b}^{a}  &  =\text{ }^{-}L_{c}^{a+}L_{b}^{c}=^{+}L_{c}^{a-}L_{b}^{c};\text{
}\\
L_{c}^{a}L_{d}^{b}\eta_{ab}  &  =\eta_{cd}\text{; }^{-}L_{c}^{a}\text{ }%
^{-}L_{d}^{b}\text{ }\eta_{ab}=\eta_{cd}\text{; }^{+}L_{c}^{a}\text{ }%
^{+}L_{d}^{b}\text{ }\eta_{ab}=\eta_{cd};\nonumber
\end{align}
and
\begin{equation}
l_{b}^{a}=\text{ }^{-}l_{b}^{a}+\text{ }^{+}l_{b}^{a},
\end{equation}
with $l_{ab}=-l_{ba}$ and similarly for $^{-}l_{ab}$ and $^{+}l_{ab}$ The
connection one-forms $\alpha_{b}^{a}$, $^{-}\alpha_{b}^{a}$ and $^{+}%
\alpha_{b}^{a}$ take values in, respectively,$l$, $^{-}l$ and $\ ^{+}l$.

Under the usual $L$ (and$^{-}L$ and $^{+}L$) - valued gauge transformations
\begin{align}
\theta^{a}  &  \rightarrow\theta_{1}^{a}=(L^{-1})_{c}^{a}\theta^{c}%
=(^{+}L^{-1})_{b}^{a}(^{-}L^{-1})_{c^{\prime}}^{b}\theta^{c},\Theta
^{a}\rightarrow(L^{-1})_{c}^{a}\Theta^{c}\\
\alpha_{b}^{a}  &  \rightarrow\alpha_{1b}^{a}=(L^{-1})_{c}^{a}dL_{b}%
^{c}+(L^{-1})_{c}^{a}\alpha_{d}^{c}L_{b}^{d}\nonumber\\
^{-}\alpha_{b}^{a}  &  \rightarrow\text{ }^{-}\alpha_{1b}^{a}=(^{-}L^{-1}%
)_{c}^{a}d\text{ }^{-}L_{b}^{c}+(^{-}L^{-1})_{c}^{a}\text{ }^{\_}\alpha
_{d}^{c}\text{ }^{-}L_{b}^{d}\nonumber\\
\digamma_{b}^{a}  &  \rightarrow\digamma_{1b}^{a}=(^{-}L^{-1})_{c}^{a}\text{
}\digamma_{d}^{c}\text{ }^{-}L_{b}^{d};\text{ }^{-}\digamma_{b}^{a}%
\rightarrow\text{ }^{-}\digamma_{1b}^{a}=(^{-}L^{-1})_{c}^{a}\text{ }%
^{-}\digamma_{d}^{c}\text{ }^{-}L_{b}^{d}.\nonumber
\end{align}
and similarly for $^{+}\alpha_{b}^{a}$ and $^{+}\digamma_{b}^{a}.$

The Cartan structure equations can be related to a connection $_{p}A$ with
values in the Lie algebra of the Poincar\'{e} group as follows. \ Let%
\begin{equation}
_{p}A=\theta^{a}V_{a}+\alpha_{b}^{a}V_{a}^{b}%
\end{equation}
where $V_{ba}=-V_{ab}$ and the generators of the Poincar\'{e} Lie algebra
satisfy the commutation relations%
\begin{align}
\lbrack V_{a},V_{b}]  &  =0,\text{ }[V_{a},V_{bc}]=\frac{1}{2}(V_{c}\eta
_{ab}-V_{b}\eta_{ac}),\\
\lbrack V_{ab},V_{cd}]  &  =\frac{1}{2}(\eta_{ac}V_{db}+\eta_{ad}V_{bc}%
+\eta_{bc}V_{ad}+\eta_{bd}V_{ca}),\nonumber
\end{align}
Then the curvature of the connection, $_{p}\digamma$ is
\begin{equation}
_{p}\digamma=d\text{ }_{p}A+\frac{1}{2}[_{p}A,\text{ }_{p}A]=\Theta^{a}%
V_{a}+\digamma_{b}^{a}V_{a}^{b},
\end{equation}
where $\Theta^{a}$ and $\digamma_{b}^{a}$ are as in Eqs.(2) and (4) above.
\ The Poincar\'{e} group is the semi-direct product of the abelian group of
translations and the Lorentz group. \ Under Poincar\'{e} group-valued gauge
transformations%
\begin{align}
\theta^{a}  &  \rightarrow\theta_{1}^{a}=(L^{-1})_{c}^{a}[\theta^{c}+D\pi
^{c}],\text{ }\Theta^{a}\rightarrow\Theta_{1}^{a}=(L^{-1})_{c}^{a}[\Theta
^{c}+\digamma_{b}^{c}\pi^{b}],\\
\text{ }\alpha_{b}^{a}  &  \rightarrow\alpha_{1b}^{a}=(L^{-1})_{c}^{a}%
dL_{b}^{c}+(L^{-1})_{c}^{a}\alpha_{d}^{c}L_{b}^{d}\nonumber
\end{align}
where $L_{b}^{a}$ is as in Eq.(8) and the function $\pi^{a}$ takes values in
the subgroup of translations $R^{4}$.

Such metric connections with non-zero torsion are related to the unique metric
and torsion-free Levi-Civita connection one-form, $\omega_{b}^{a}$, as
follows. \ If the components of $\alpha_{b}^{a}$, $\omega_{b}^{a}$ and
$\Theta^{a}$ are given by%
\begin{equation}
\alpha_{b}^{a}=\alpha_{bc}^{a}\theta^{c},\text{ }\omega_{b}^{a}=\omega
_{bc}^{a}\theta^{c},\text{ }\Theta^{a}=\frac{1}{2}\Theta_{bc}^{a}\theta
^{b}\theta^{c}%
\end{equation}
where $\Theta_{bc}^{a}=-\Theta_{cb}^{a}$ then
\begin{equation}
\omega_{ab}=\alpha_{ab}+\sigma_{ab},
\end{equation}
where%
\begin{equation}
\sigma_{ab}=\sigma_{abc}\theta^{c}=\frac{1}{2}(\Theta_{abc}-\Theta
_{bac}-\Theta_{cab})\theta^{c},
\end{equation}
and conversely%
\begin{equation}
\Theta_{abc}=\mathbf{\sigma}_{abc}-\sigma_{acb}.
\end{equation}

When they are expressed in terms of the Levi-Civita connection the first
Cartan structure equations become%
\begin{equation}
D\theta^{a}\equiv d\theta^{a}+\omega_{b}^{a}\theta^{b}=d\theta^{a}+\text{
}^{-}\omega_{b}^{a}\theta^{b}+\text{ }^{+}\omega_{b}^{a}\theta^{b}=0
\end{equation}
where $D$ is now the exterior covariant derivative with respect to the
Levi-Civita connection $\omega_{b}^{a}$. \ Here $^{-}\omega_{b}^{a}$ and
$^{+}\omega_{b}^{a}$ are respectively the anti self-dual and self dual parts
of the,\ uniquely torsion-free and metric, Levi-Civita connection. \ The
second Cartan structure equations are then%
\begin{equation}
\Omega_{b}^{a}=d\text{ }\omega_{b}^{a}+\text{ }\omega_{c}^{a}\text{ }%
\omega_{b}^{c}=\frac{1}{2}R_{bcd}^{a}\theta^{c}\theta^{d}%
\end{equation}
where $\Omega_{b}^{a}$ denotes the curvature two-form with Riemann tensor
components $R_{bcd}^{a}$. \ In terms of their anti self-dual and self-dual
parts $\Omega_{b}^{a}=^{-}\Omega_{b}^{a}+^{+}\Omega_{b}^{a}$, $^{-}\Omega
_{b}^{a}=d$ $^{-}\omega_{b}^{a}+$ $^{-}\omega_{c}^{a}$ $^{-}\omega_{b}%
^{c}=\frac{1}{2}^{-}R_{bcd}^{a}\theta^{c}\theta^{d}$ and similarly for
$^{+}\Omega_{b}^{a}$.

The first and second Bianchi identities are now%
\begin{align}
\Omega_{b}^{a}\theta^{a}  &  =\ ^{-}\Omega_{b}^{a}\theta^{b}+\text{ }%
^{+}\Omega_{b}^{a}\theta^{b}=0,\\
D\Omega_{b}^{a}  &  =\text{ }^{-}D^{-}\Omega_{b}^{a}=\text{ }^{+}D^{+}%
\Omega_{b}^{a}=0,\nonumber
\end{align}
where now $D,$ $^{-}D$ and $^{+}D$ respectively denote the relevant exterior
covariant derivative with respect to the Levi-Civita connection and the
antiself-dual or self-dual part of the connection. \ In four space-time
dimensions the second Cartan equations imply that $D(^{\ast}\Omega_{b}^{a})=0$
where the dual curvature two-form is $^{\ast}\Omega_{b}^{a}=\frac{1}%
{2}\epsilon_{bcd}^{a}\Omega^{cd}.$ \ The Ricci tensor has components
$R_{ab}=R_{.acb}^{c}.$

As zero curvature connections will be important subsequently recall that the
curvature of the connection $\alpha_{b}^{a}$ vanishes if and only if
$\alpha_{b}^{a}=(L^{-1})_{c}^{a}dL_{b}^{c}$ where $L_{b}^{a}$ is a
$SO(1,3)-$valued function. \ Furthermore, in the case of a flat Levi-Civita
connection, local coordinates, $x^{b},$ can be chosen so that%
\begin{align}
\theta^{a}  &  =(L^{-1})_{b}^{a}dx^{b},\\
\omega_{b}^{a}  &  =(L^{-1})_{c}^{a}dL_{b}^{c}\nonumber
\end{align}
When the curvature of the Levi-Civita connection vanishes in addition to the
flat $so(1,3)$-valued connection $\omega_{b}^{a}=(L^{-1})_{c}^{a}dL_{b}^{c}$
the pair $\{\theta^{a}$, $\omega_{b}^{a}\}$ determine a flat connection,
$_{p}A_{flat}$, that has values in the Lie algebra of the Poincar\'{e}
(inhomogeneous Lorentz) group. \ From above%
\begin{equation}
_{p}A_{flat}=(L^{-1})_{b}^{a}dx^{b}V_{a}+(L^{-1})_{c}^{a}dL_{b}^{c}V_{a}^{b}.
\end{equation}
A $5\times5$ matrix representation of such a flat connection can be
constructed by using the $5\times5$ matrix representation of a Poincar\'{e}
group-valued function $g$%
\begin{equation}
g=[g_{\beta}^{\alpha}]=\left[
\begin{array}
[c]{cc}%
L_{b}^{a} & x^{a}\\
0 & 1
\end{array}
\right]
\end{equation}
($\alpha$, $\beta$ range and sum over 1 to 5). \ This representation is then
given by $_{p}A_{flat}=g^{-1}dg.$

Einstein's gravitational field equations, expressed in terms of components
with respect to the co-frame $\theta^{a}$ and Levi-Civita connection
$\omega_{b}^{a}$, are%

\begin{equation}
G_{ab}=\Lambda\eta_{ab}+8\pi\kappa T_{ab}%
\end{equation}
and can be expressed as%
\begin{equation}
^{\ast}\Omega_{ab}\theta^{b}=-G_{ba}\eta^{a}=-(\Lambda\eta_{ba}+8\pi\kappa
T_{ba})\eta^{b},
\end{equation}
where $G_{ba}$ is the Einstein tensor, $G_{ba}=R_{ba}-\frac{1}{2}R\eta_{ba}$,
or equivalently as%
\begin{equation}
^{-}\Omega_{ab}\theta^{b}=-\frac{i}{2}G_{ba}\eta^{a}=-\frac{i}{2}(\Lambda
\eta_{ba}+8\pi\kappa T_{ba})\eta^{b}%
\end{equation}
(and its complex conjugate). \ Here $\Lambda$ denotes the cosmological
constant, $T_{ba}$ are the components of the matter energy-momentum tensor and
$\kappa$ is the coupling constant and the three-forms $\eta^{a}$ are given by
$\eta^{a}\equiv\frac{1}{3!}\epsilon_{bcd}^{a}\theta^{b}\theta^{c}\theta^{d}$
or equivalently $\theta^{a}\theta^{b}\theta^{c}=\epsilon^{abcd}\eta_{d}$.

Subsequently, the vacuum field equations, with $\Lambda=T_{ba}=0,$ will, in
the main, be considered first then the non-vacuum equations will be discussed
in a similar but less detailed way.

\section{Type N=2 generalized forms}

A\ type $N=2$ generalized differential $p-$ form on a $n$-dimensional manifold
$M$ can be expressed as an expansion in terms of ordinary forms and two
linearly independent forms of degree minus one $\mathbf{m}^{i}$ ($i,j.k..$
range and sum from $1$ to $2$). \ Type $N=2$ forms include two special cases,
type $N=0$ generalized forms which are ordinary forms and type $N=1$ forms
which are are expressed in terms of ordinary forms and only one minus one
form, $\mathbf{m}^{1}$ \cite{rob5}, \cite{rob6}.

For a given choice of minus one-forms, $\mathbf{m}^{i}$ , a type $N=2$
generalized p-form, $\overset{p}{\mathbf{r}}$, can be written as%
\begin{equation}
\overset{p}{\mathbf{r}}=\overset{p}{\rho}+\overset{p+1}{\rho}_{i}%
\mathbf{m}^{i}+\overset{p+2}{\rho}\mathbf{m}^{1}\mathbf{m}^{2},
\end{equation}
where $-2\leqq p\leqq n$ and the ordinary forms $\overset{p}{\rho
},\overset{p+1}{\rho}_{i},\overset{p+2}{\rho}$ are respectively a $p-$form,
two ($p+1)-$forms and a ($p+2)-$form. \ Hence for a given pair, $\mathbf{m}%
^{i},$ the generalized form is determined by four ordinary differential forms
and is zero if and only if all those ordinary forms vanish. \ Such generalized
\ forms obey the same basic rules of exterior algebra as ordinary differential
forms. \ The two independent degree minus one forms\textbf{ }$\mathbf{m}^{i}$
satisfy the ordinary distributive and associative laws of exterior algebra and
the exterior product rule so that%
\begin{equation}
\overset{p}{\rho}\mathbf{m}^{i}=(-1)^{p}\mathbf{m}^{i}\overset{p}{\rho};\text{
}\mathbf{m}^{i}\mathbf{m}^{j}=-\mathbf{m}^{j}\mathbf{m}^{i},\text{ }%
\mathbf{m}^{1}\mathbf{m}^{2}\neq0.
\end{equation}
If $\overset{p}{\mathbf{r}}$and $\overset{q}{\mathbf{s}}$ are respectively
generalized $p-$ and $q-$forms, then $\overset{p}{\mathbf{r}}%
\overset{q}{\mathbf{s}}=(-1)^{pq}\overset{q}{\mathbf{s}}\overset{p}{\mathbf{r}%
}$.

It is assumed that the exterior derivative, $\mathbf{d},$ of generalized forms
increases the degree of a form by one, agrees with the usual exterior
derivative when acting on ordinary forms, and satisfies the usual properties,
in particular\
\begin{align}
\mathbf{d}(\overset{p}{\mathbf{r}}\overset{q}{\mathbf{s}})  &  =(\mathbf{d}%
\overset{p}{\mathbf{r}})\overset{q}{\mathbf{s}}+(-1)^{p}\overset{p}{\mathbf{r}%
}(\mathbf{d}\overset{q}{\mathbf{s}}),\\
\mathbf{d}^{2}  &  =0,\nonumber
\end{align}
It follows that the exterior derivatives of the minus one-forms $\mathbf{m}%
^{i}$ are type $N=2$ generalized zero-forms and they can be chosen, as will be
done in this paper, so that%
\begin{equation}
\mathbf{dm}^{1}=1\text{ and }\mathbf{dm}^{2}=0.
\end{equation}
If $\varphi$ is a smooth map between manifolds $P$ and $M,$ $\varphi
:P\rightarrow M,$ then the induced map of type $N=2$ generalized forms,
$\varphi_{(2)}^{\ast}$, is the linear map defined by using the standard
pull-back map, $\varphi^{\ast}$, for ordinary form%
\begin{equation}
\varphi_{(2)}^{\ast}(\overset{p}{\mathbf{r}})=\varphi^{\ast}(\overset{p}{\rho
})+\varphi^{\ast}(\overset{p+1}{\rho}_{i})\mathbf{m}^{i}+\varphi^{\ast
}(\overset{p+2}{\rho})\mathbf{m}^{1}\mathbf{m}^{2},
\end{equation}
and $\varphi_{(2)}^{\ast}(\overset{p}{\mathbf{r}}\overset{q}{\mathbf{s}%
})=\varphi_{(2)}^{\ast}(\overset{p}{\mathbf{r}})\varphi_{(2)}^{\ast
}(\overset{q}{\mathbf{s}})$. \ Hence $\varphi_{(2)}^{\ast}(\mathbf{m}%
^{i})=\mathbf{m}^{i}$.

In addition to the representation used here generalized forms can also be
represented by using superspace \cite{rob7}

All closed generalized forms can be written explicitly as exact forms. The
generalized $p-$form $\overset{p}{\mathbf{r}}$ in Eq.(28) is closed if and
only if%
\begin{align}
\overset{p}{\mathbf{r}}  &  =\overset{p}{\rho}+(-1)^{p}d\overset{p}{\rho
}\mathbf{m}^{1}+\overset{p+1}{\rho}_{2}\mathbf{m}^{2}+(-1)^{p+1}%
d\overset{p+1}{\rho}_{2}\mathbf{m}^{1}\mathbf{m}^{2}\\
&  =\mathbf{d[}(-1)^{p}\overset{p}{\rho}\mathbf{m}^{1}+(-1)^{p+1}%
\overset{p+1}{\rho}_{2}\mathbf{m}^{1}\mathbf{m}^{2}+\mathbf{d}%
\overset{p-2}{\mathbf{s}}]\nonumber
\end{align}
where $\overset{p-2}{\mathbf{s}}$ is any generalized $(p-2)-$%
form\footnote{Generalized forms can be related to relative cohomology classes
and the Thom class of a vector bundle \cite{pop}, \cite{bott}. $\ $Let $S$ be
a submanifold of $M$ and let $i$ be the inclusion map $i:$ $S\rightarrow M$.
\ Then if $\chi$ is a closed $p-$form which is not exact on $M$ but which
pulls back to an exact form on $S$ with $i^{\ast}(\chi)=d\varsigma$, for
some\ ($p-1)-$form $\varsigma$, in other words if $\mathbf{p=}(-1)^{p-1}%
\varsigma+i^{\ast}(\chi)\mathbf{m}$ is a closed, and hence exact type $N=1$
form on $S,$ then $\chi$ represents a cohomology class in the relative de Rham
cohomology $H^{p}(M,S)$.}.

Integration is defined using polychains, \cite{rob6}. \ A $p-$polychain of
type $N=2$ in $M$, denoted $\mathbf{c}_{p}$ is an ordered quadruple of
ordinary (real, singular) chains in $M$%
\begin{equation}
\mathbf{c}_{p}=(c_{p},c_{p+1}^{1},c_{p+1}^{2},c_{p+2}),
\end{equation}
where $c_{p}$ is an ordinary $p-$chain, $c_{p+1}^{1}$ and $c_{p+1}^{2}$ are
ordinary $p+1-$chains and $c_{p+2}$ is and ordinary ordinary\ $p+2-$chain.
\ The ordinary chains' boundaries are denoted by $\partial$ and the boundary
of the polychain $\mathbf{c}_{p}$ is the $(p-1)-$polychain $\mathbf{\partial
c}_{p}$ given by%
\begin{equation}
\mathbf{\partial c}_{p}=(\partial c_{p},\partial c_{p+1}^{1}+(-1)^{p}%
c_{p},\partial c_{p+1}^{2},\partial c_{p+2}+(-1)^{p+1}c_{p+1}^{2}),
\end{equation}
and%
\begin{equation}
\partial^{2}\mathbf{c}_{p}=0.
\end{equation}

When $N=2$ the integral of a generalized form $\overset{p}{\mathbf{r}}$ over a
polychain\textbf{ }$\mathbf{c}_{p}$ is%
\begin{equation}
\int_{\mathbf{c}_{p}}\overset{p}{\mathbf{r}}=\int_{c_{p}}\overset{p}{\rho
}+\int_{c_{p+1}^{1}}\overset{p+1}{\rho}_{1}+\int_{c_{p+1}^{2}}%
\overset{p+1}{\rho}_{2}+\int_{c_{p+2}}\overset{p+2}{\rho}.
\end{equation}
and Stokes' theorem applies%
\begin{equation}
\int_{\mathbf{c}_{p}}d\overset{p-1}{\mathbf{r}}=\int_{\partial\mathbf{c}_{p}%
}\overset{p-1}{\mathbf{r}}.
\end{equation}

\section{Generalized connections and Einstein's field equations in four
dimensional space-times}

\ Consider now the generalization of the ordinary first Cartan structure
equations with non-zero torsion given by%
\begin{equation}
\mathbf{D}\theta^{a}\equiv d\theta^{a}+\mathbf{A}_{b}^{a}\theta^{b}%
=\mathbf{T}^{a},
\end{equation}
where the generalized connection in these equations, here and subsequently, is
chosen to be the metric connection $\mathbf{A}_{b}^{a}$, with generalized
curvature $\mathbf{F}_{b}^{a}$, given by%
\begin{align}
\mathbf{A}_{b}^{a}  &  =\alpha_{b}^{a}-\digamma_{b}^{a}\mathbf{m}^{1}+\text{
}^{\ast}\digamma_{b}^{a}\mathbf{m}^{2},\\
\mathbf{F}_{b}^{a}  &  =d\alpha_{b}^{a}+\alpha_{c}^{a}\alpha_{b}^{c}%
-\digamma_{b}^{a}-D\digamma_{b}^{a}\mathbf{m}^{1}+D(\text{ }^{\ast}%
\digamma_{b}^{a})\mathbf{m}^{2},\nonumber
\end{align}
where $\mathbf{D}$ is the exterior covariant derivative with respect to
$\mathbf{A}_{b}^{a}$ and $\alpha_{b}^{a}$, $\digamma_{b}^{a}$ and its dual
$^{\ast}\digamma_{b}^{a}$ (as in Section 2) are respectively an ordinary
connection one-form with exterior covariant derivative $D$ and two-forms with
values in the Lie algebra of the Lorentz group. \ The generalized curvature
$\mathbf{F}_{b}^{a}$ is zero if and only if $\digamma_{b}^{a}$ is the
curvature of $\alpha_{b}^{a}$.

If the generalized torsion two-form is chosen to be non-zero and equal to%
\begin{equation}
\mathbf{T}^{a}=(\Lambda\delta_{b}^{a}+8\pi\kappa T_{b}^{a})\eta^{b}%
\mathbf{m}^{2}%
\end{equation}
Then Eqs.(39-41) hold if and only if $\alpha_{b}^{a}$ is the Levi-Civita
connection, $\omega_{b}^{a},$ for the metric $ds^{2}=\eta_{ab}\theta
^{a}\otimes\theta^{b}$, $\digamma_{b}^{a}$ is its curvature two-form
$\Omega_{b}^{a}$ and the metric satisfies Einstein's field equations, Eq.(25),
with cosmological constant $\Lambda$ and matter energy momentum tensor
$T_{b}^{a}$. \ Furthermore in this case the generalized curvature of
$\mathbf{A}_{b}^{a}$ is zero.

\ In order to further explore the equations above it is convenient to
introduce, and at times to use, as an alternative to the pair $\mathbf{m}%
^{1},$ $\mathbf{m}^{2}$, a basis of minus one-forms given by complex conjugate
minus one-forms, $\mathbf{m}$ and $\overline{\mathbf{m}},$%
\begin{equation}
\mathbf{m=}(\mathbf{m}^{1}+i\mathbf{m}^{2}),\text{ }\overline{\mathbf{m}%
}=(\mathbf{m}^{1}-i\mathbf{m}^{2}),
\end{equation}
for then the complex conjugate self-dual and anti-self dual parts of
$\mathbf{A}_{b}^{a}$ and its curvature $\mathbf{F}_{b}^{a}$ can then be simply
expressed and separated, and $\mathbf{A}_{b}^{a}$ written in terms of its anti
self-dual and self dual parts is%
\begin{align}
\mathbf{A}_{b}^{a}  &  =\alpha_{b}^{a}-^{-}\digamma_{b}^{a}\mathbf{m-}%
^{+}\digamma_{b}^{a}\overline{\mathbf{m}}\text{ }=\text{ }^{-}\mathbf{A}%
_{b}^{a}+\text{ }^{+}\mathbf{A}_{b}^{a},\\
\mathbf{F}_{b}^{a}  &  =\mathbf{=}d\alpha_{b}^{a}+\alpha_{c}^{a}\alpha_{b}%
^{c}-(^{-}\digamma_{b}^{a}+\text{ }^{+}\digamma_{b}^{a})-\text{ }^{-}%
D^{-}\digamma_{b}^{a}\mathbf{m}\nonumber\\
&  -\text{ }^{+}D\text{ }^{+}\digamma_{b}^{a}\overline{\mathbf{m}}.\nonumber
\end{align}
Eq.(39) then becomes%
\begin{equation}
\mathbf{D}\theta^{a}\equiv D\theta^{a}+\text{ }^{-}\digamma_{b}^{a}\theta
^{b}\mathbf{m}+\text{ }^{+}\digamma_{b}^{a}\theta^{b}\overline{\mathbf{m}%
}=\mathbf{T}^{a}.
\end{equation}
The complex conjugate generalized connection one-forms $^{-}\mathbf{A}_{b}%
^{a}$ and $^{+}\mathbf{A}_{b}^{a}$ are given by%
\begin{align}
^{-}\mathbf{A}_{b}^{a}  &  =\text{ }^{-}\alpha_{b}^{a}-^{-}\digamma_{b}%
^{a}\mathbf{m},\\
^{+}\mathbf{A}_{b}^{a}  &  =\text{ }^{+}\alpha_{b}^{a}-^{+}\digamma_{b}%
^{a}\overline{\mathbf{m}}.\nonumber
\end{align}
and the generalized curvature two-forms of these connections are
\begin{align}
^{-}\mathbf{F}_{b}^{a}  &  =d\text{ }^{-}\mathbf{A}_{b}^{a}+\text{ }%
^{-}\mathbf{A}_{c}^{a}\text{ }^{-}\mathbf{A}_{b}^{c}\\
&  =d\text{ }^{-}\alpha_{b}^{a}+\text{ }^{-}\alpha_{c}^{a}\text{ }^{-}%
\alpha_{b}^{c}-\text{ }^{-}\digamma_{b}^{a}-\text{ }^{-}D^{-}\digamma_{b}%
^{a}\mathbf{m}\nonumber
\end{align}
with $^{+}\mathbf{F}_{b}^{a}$ being the complex conjugate of $^{-}%
\mathbf{F}_{b}^{a}$.

Whenever $\digamma_{b}^{a}$ is the curvature of the ordinary connection
$\alpha_{b}^{a}$ these generalized curvatures are all zero, irrespective of
field equations, and all these generalized connections are flat. \ In this
case%
\begin{align}
^{-}\mathbf{A}_{b}^{a}  &  =(^{-}\mathbf{L}^{-1})_{c}^{a}\mathbf{d}%
(^{-}\mathbf{L)}_{b}^{c}\text{ where }^{-}\mathbf{L}_{_{b}}^{a}=\delta_{b}%
^{a}-^{-}\alpha_{b}^{a}\mathbf{m},\\
^{+}\mathbf{A}_{b}^{a}  &  =(^{+}\mathbf{L}^{-1})_{c}^{a}\mathbf{d}%
(^{+}\mathbf{L)}_{b}^{c}\text{ where }^{+}\mathbf{L}_{_{b}}^{a}=\delta_{b}%
^{a}-^{+}\alpha_{b}^{a}\overline{\mathbf{m}}\mathbf{,}\nonumber\\
\mathbf{A}_{b}^{a}  &  =(\mathbf{L}^{-1})_{c}^{a}\mathbf{dL}_{b}^{c}\text{
where }\nonumber\\
\mathbf{L}_{_{b}}^{a}  &  =^{-}\mathbf{L}_{_{c}}^{a+}\mathbf{L}_{b}^{c}%
=^{+}\mathbf{L}_{_{c}}^{a-}\mathbf{L}_{b}^{c},\nonumber
\end{align}
and $\eta_{ab}\mathbf{L}_{c}^{a}\mathbf{L}_{d}^{b}=\eta_{cd}$ and similarly
for$^{-}\mathbf{L}_{_{b}}^{a}$ and $^{+}\mathbf{L}_{_{b}}^{a}$.

When $\mathbf{A}_{b}^{a}$ is flat it follows from Eq.(39) that for any
generalized torsion two-form%
\begin{equation}
\mathbf{DT}^{a}=0
\end{equation}
or equivalently\footnote{Eq. (49) implies that $\mathbf{d}$($\mathbf{L}%
_{b}^{a}\mathbf{T}^{b})=0$ and in the general relativistic case, with
$\mathbf{T}^{b}$ given by Eq.(41), this is the usual local conservation law
$\nabla_{a}T^{ab}=0$.}%
\begin{equation}
\mathbf{d(L}_{b}^{a}\theta^{b})=\mathbf{L}_{b}^{a}\mathbf{T}^{b}.
\end{equation}
\ Hence the generalized one-forms $\mathbf{L}_{b}^{a}\theta^{b}$ are closed,
and therefore exact, if and only if\textbf{ }the generalized torsion is zero,
that is when%

\begin{equation}
\mathbf{D}\theta^{a}\equiv d\theta^{a}+\mathbf{A}_{b}^{a}\theta^{b}%
=d\theta^{a}+\text{ }^{-}\mathbf{A}_{b}^{a}\theta^{b}+\text{ }^{+}%
\mathbf{A}_{b}^{a}\theta^{b}=\mathbf{0}.
\end{equation}
Now consider Eq.(44) in two cases. \ first in the case where the generalized
torsion is zero and Einstein's vacuum field equations are satisfied and second
in the case where it is non-zero and Einstein's equations with non-zero
cosmological constant and/or energy-momentum tensor are satisfied.

In the first case it follows from Eq.(50) and Eq.(49), that the co-frame
$\theta^{a}$ can be expressed as
\begin{equation}
\theta^{a}=(\mathbf{L}^{-1})_{b}^{a}\mathbf{q}^{b}=(^{+}\mathbf{L}%
^{-1}\mathbf{)}_{c}^{a}\text{ (}^{-}\mathbf{L}^{-1}\mathbf{)}_{b}%
^{c})\mathbf{q}^{b}%
\end{equation}
where $\mathbf{q}^{b}$ are four independent closed generalized one-forms.
\ All closed type \ $N\geqq1$ generalized forms are exact. \ Hence the closed
one-forms $\mathbf{q}^{b}$ are exact, that is $\mathbf{q}^{b}=\mathbf{dx}^{b}%
$. \ Consequently a co-frame for any vacuum solution is given by%
\begin{equation}
\theta^{a}=(\mathbf{L}^{-1})_{b}^{a}\mathbf{dx}^{b}.
\end{equation}
The generalized zero-forms $\mathbf{x}^{b}$ are unique, modulo $\mathbf{du}%
^{b}$ where $\mathbf{u}^{b}$ are any generalized minus one-forms.

For any co-frame $\theta^{a}$, connection $\alpha_{b}^{a},$ and $\mathbf{L}%
_{b}^{a}$ given by Eq.(47), Eq.(52) can be solved for the generalized
zero-forms $\mathbf{x}^{b}$ and their exterior derivatives. \ The solutions
are%
\begin{align}
\mathbf{x}^{a}  &  =-\frac{1}{2}\theta^{a}\mathbf{m-}\frac{1}{2}\theta
^{a}\overline{\mathbf{m}}+\frac{1}{2}(^{+}\alpha_{b}^{a}\theta^{b}-\text{
}^{-}\alpha_{b}^{a}\theta^{b})\mathbf{m}\overline{\mathbf{m}}+\mathbf{du}%
^{a},\\
\mathbf{dx}^{a}  &  =\theta^{a}+\text{ }^{-}\alpha_{b}^{a}\theta
^{b}\mathbf{m+}^{+}\alpha_{b}^{a}\theta^{b}\overline{\mathbf{m}}+\frac{1}%
{2}d(^{+}\alpha_{b}^{a}\theta^{b}-\text{ }^{\_}\alpha_{b}^{a}\theta
^{b})\mathbf{m}\overline{\mathbf{m}}-\frac{1}{2}D\theta^{a}(\mathbf{m+}%
\overline{\mathbf{m}}\mathbf{)}.\nonumber
\end{align}
and then%
\begin{align}
(\mathbf{L}^{-1})_{b}^{a}\mathbf{dx}^{b}  &  =\theta^{a}+\frac{i}%
{2}\mathcal{F}_{b}^{a}\eta^{b}\mathbf{m}\overline{\mathbf{m}}-\frac{1}%
{2}D\theta^{a}(\mathbf{m+}\overline{\mathbf{m}}\mathbf{)},\\
\mathcal{F}_{b}^{a}  &  =\digamma_{\text{ }bc}^{c}\text{.}^{a}-\frac{1}%
{2}\digamma_{\text{ \ }cd}^{cd}\delta_{b}^{a}.\nonumber
\end{align}

In particular Eq.(50), and hence Eqs.(51 \& 52), is solved for $\mathbf{x}%
^{b}$ by Eq.(53) if and only if $D\theta^{a}=0$ so that that $\alpha_{b\text{
}}^{a}$ is the Levi-Civita connection $\omega_{b}^{a}$ corresponding to the
orthonormal co-frame $\theta^{a}$, and $\digamma_{b}^{a}=\Omega_{b}^{a}$ so
that $\mathcal{F}_{b}^{a}=G_{b}^{a}=0$.

\textbf{In summary:} \ The four generalized one-forms $(\mathbf{L}^{-1}%
)_{b}^{a}\mathbf{dx}^{b}$ defined by Eqs.(47, 52 \& 53) determine a solution
of Einstein's vacuum field equations if and only if they are linearly
independent and equal to a co-frame of ordinary one-forms. \ Einstein's vacuum
field equations are then satisfied by the metric $ds^{2}=\eta_{ab}\theta
^{a}\otimes\theta^{b}$ where the ordinary co-frame is $\theta^{a}$ and
$\alpha_{b}^{a}$ is the corresponding Levi-Civita connection $\omega_{b}^{a}$
so that
\begin{align}
\theta^{a}  &  =(\mathbf{L}^{-1})_{b}^{a}\mathbf{dx}^{b},\\
\mathbf{x}^{a}  &  =-\frac{1}{2}\theta^{a}\mathbf{m-}\frac{1}{2}\theta
^{a}\overline{\mathbf{m}}+\frac{1}{2}(^{+}\omega_{b}^{a}\theta^{b}-\text{
}^{-}\omega_{b}^{a}\theta^{b})\mathbf{m}\overline{\mathbf{m}}+\mathbf{du}%
^{b},\nonumber\\
\mathbf{dx}^{a}  &  =\theta^{a}+\text{ }^{-}\omega_{b}^{a}\theta
^{b}\mathbf{m+}^{+}\omega_{b}^{a}\theta^{b}\overline{\mathbf{m}}+\frac{1}%
{2}d(^{+}\omega_{b}^{a}\theta^{b}-\text{ }^{\_}\omega_{b}^{a}\theta
^{b})\mathbf{m}\overline{\mathbf{m}},\nonumber
\end{align}

The expression, $(\mathbf{L}^{-1})_{b}^{a}\mathbf{dx}^{b}$, for any co-frame
determining a vacuum solution of Einstein's equations looks formally similar
to the Minkowski space-time expression, Eq.(22), with generalized zero-forms
replacing ordinary zero-forms. \ However the generalized coordinates
$\mathbf{x}^{b}$ are constrained by the condition that $(\mathbf{L}%
^{-1}\mathbf{)}_{b}^{a}\mathbf{dx}^{b}$ equals a co-frame of ordinary forms
$\theta^{a}$, that is that the zero-forms $\mathbf{x}^{b}$ are given by Eq.(55).

Consider now the second case where\ the generalized torsion $\mathbf{T}^{a}$
is non-zero. \ The generalized connection one-form $\mathbf{A}_{b}^{a}$ in
Eq.(44), with $\digamma_{b}^{a}$ the curvature of $\alpha_{b}^{a},$ can be
replaced by a second (metric) generalized connection one-form,
$\widetilde{\mathbf{A}}$ $_{b}^{a}$ which also has values in the Lie algebra
of the Lorentz group but which has zero generalized torsion. \ This is done by
following same procedure as the one outlined for ordinary forms in Section 2,
and by using the expansions in terms of components given by%
\begin{align}
\widetilde{\mathbf{A}}_{b}^{a}  &  =\widetilde{\mathbf{A}}_{bc}^{a}\theta
^{c},\text{ }\mathbf{A}_{b}^{a}=\mathbf{A}_{bc}^{a}\theta^{c},\\
\mathbf{T}^{a}  &  =\frac{1}{2}\mathbf{T}_{bc}^{a}\theta^{b}\theta
^{c}.\nonumber
\end{align}
Then%
\begin{equation}
\widetilde{\mathbf{A}}_{ab}=\mathbf{A}_{ab}+\mathbf{S}_{ab},
\end{equation}
where%
\begin{equation}
\mathbf{S}_{ab}=\mathbf{S}_{abc}\theta^{c}=\frac{1}{2}(\mathbf{T}%
_{abc}-\mathbf{T}_{bac}-\mathbf{T}_{cab})\theta^{c},
\end{equation}
and conversely%
\begin{equation}
\mathbf{T}_{abc}=\mathbf{S}_{abc}-\mathbf{S}_{acb}%
\end{equation}
For the generalized torsion in Eq.(41) the components are%
\begin{equation}
\mathbf{T}_{bc}^{a}=\frac{1}{3}(\Lambda\delta_{e}^{a}+8\pi\kappa T_{e}%
^{a})\epsilon_{bcd}^{e}\theta^{d}\mathbf{m}^{2},
\end{equation}
and $\mathbf{S}_{ab}$ is%
\begin{equation}
\mathbf{S}_{ab}=\frac{1}{6}[\Lambda\epsilon_{abcd}+8\pi\kappa(T_{ae}%
\epsilon_{bcd}^{e}-T_{be}\epsilon_{acd}^{e})+4\pi\kappa(T_{de}\epsilon
_{abc}^{e}-T_{ce}\epsilon_{abd}^{e})]\theta^{c}\theta^{d}\mathbf{m}^{2}.
\end{equation}
With this choice Einstein's equations, Eq.(25), are equivalent to the Cartan
equations%
\begin{equation}
\widetilde{\mathbf{D}}\theta^{a}\equiv\mathbf{d}\theta^{a}%
+\widetilde{\mathbf{A}}_{b}^{a}\theta^{b}=0.
\end{equation}
where $\widetilde{\mathbf{D}}$ denotes the exterior covariant derivative with
respect to $\widetilde{\mathbf{A}}_{b}^{a}$. \ The curvature of the connection
$\widetilde{\mathbf{A}}_{b}^{a}$ in Eq.(66) is
\begin{align}
\widetilde{\mathbf{F}}_{b}^{a}  &  =\mathbf{d}\widetilde{\mathbf{A}}_{b}%
^{a}+\widetilde{\mathbf{A}}_{c}^{a}\widetilde{\mathbf{A}}_{b}^{c}\\
&  =\mathbf{DS}_{b}^{a},\nonumber
\end{align}
where $\mathbf{D}$ is, again, the exterior covariant derivative with respect
to the flat connection $\mathbf{A}_{b}^{a}$.

As an illustration of these results consider the case where the
energy-momentum tensor is zero but the cosmological constant is non-zero.

When the matter energy-momentum tensor is zero but the cosmological constant
is non-zero it follows from Eqs.(39-41) that Einstein's equations,
$G_{ab}=\Lambda\eta_{ab},$ are equivalent to the equations
\begin{equation}
\mathbf{D}\theta^{a}=\mathbf{T}^{a}=\Lambda\eta^{a}\mathbf{m}^{2}.
\end{equation}
Now, from Eqs.(57 \& 61)
\begin{align}
\mathbf{S}_{ab}  &  =\frac{1}{3}\Lambda^{\ast}\Sigma_{ab}\mathbf{m}^{2},\\
\widetilde{\mathbf{A}}_{b}^{a}  &  =\mathbf{A}_{b}^{a}+\frac{1}{3}%
\Lambda^{\ast}\Sigma_{b}^{a}\mathbf{m}^{2},\nonumber
\end{align}
where $\Sigma_{b}^{a}=\theta^{a}\theta_{b}$ and $^{\ast}\Sigma_{b}^{a}%
=\frac{1}{2}\epsilon_{bcd}^{a}\theta^{c}\theta^{d}$.

Hence Einstein's equations for the metric $ds^{2}=\eta_{ab}\theta^{a}%
\otimes\theta^{b}$ with Levi-Civita connection $\omega_{b}^{a},$ and
\ non-zero cosmological constant $\Lambda$ are equivalent to the generalized
Cartan structure equations
\begin{equation}
\widetilde{\mathbf{D}}\theta^{a}\equiv d\theta^{a}+\mathbf{(A}_{b}^{a}%
+\frac{1}{3}\Lambda^{\ast}\Sigma_{b}^{a}\mathbf{m}^{2})\theta^{b}=0.
\end{equation}
Calculation of the generalized curvature, $\widetilde{\mathbf{F}}$ of the
connection $\widetilde{\mathbf{A}}_{b}^{a}$ given by Eq.(65) shows that it is
zero. \ Hence the generalized connection $\widetilde{\mathbf{A}}_{b}^{a}$ is
also flat and is%
\begin{align}
\widetilde{\mathbf{A}}_{b}^{a}  &  =\mathbf{A}_{b}^{a}+\frac{i\Lambda}%
{6}\text{ }^{\ast}\Sigma_{b}^{a}(\overline{\mathbf{m}}-\mathbf{m)}=\text{
}^{-}\widetilde{\mathbf{A}}_{b}^{a}+\text{ }^{+}\widetilde{\mathbf{A}}_{b}%
^{a},\text{ }\\
^{-}\mathbf{A}  &  =\text{ }^{-}\omega_{b}^{a}-[^{-}\Omega_{b}^{a}%
+\frac{\Lambda}{6}\text{ }^{-}\Sigma_{b}^{a}]\mathbf{m}+\frac{\Lambda}%
{6}\text{ }^{-}\Sigma_{b}^{a}\overline{\mathbf{m}}\nonumber\\
&  =\text{ }^{-}\omega_{b}^{a}-\text{ }^{-}\Omega_{b}^{a}\mathbf{m}^{1}%
-i[^{-}\Omega_{b}^{a}+\frac{\Lambda}{3}\text{ }^{-}\Sigma_{b}^{a}%
]\mathbf{m}^{2}%
\end{align}
where the anti self-dual and the (complex conjugate) self-dual parts of
$\widetilde{\mathbf{A}}_{b}^{a}$ and $\Sigma_{b}^{a}$ are denoted by
superscripts $-$ and $+$ as before.

By following an analogous procedure to the one used in the vacuum case the
following similar expressions for $\widetilde{\mathbf{A}}_{b}^{a}$ and
$\theta^{a}$ can be obtained in terms of generalized forms
$\widetilde{\mathbf{L}}_{c}^{a},$ $^{-}\widetilde{\mathbf{L}}_{c}^{a}%
,^{+}\widetilde{\mathbf{L}}_{c}^{a}$ satisfying $\eta_{ab}%
\widetilde{\mathbf{L}}_{c}^{a}\widetilde{\mathbf{L}}_{d}^{b}=\eta_{cd}$ (and
similarly for $^{-}\widetilde{\mathbf{L}}_{c}^{a}$ and $^{+}%
\widetilde{\mathbf{L}}_{c}^{a})$%
\begin{align}
\widetilde{\mathbf{A}}_{b}^{a}  &  =(\widetilde{\mathbf{L}}^{-1})_{c}%
^{a}\mathbf{d}\widetilde{\mathbf{L}}_{b}^{c},\\
\widetilde{\mathbf{A}}_{b}^{a}  &  =(^{-}\widetilde{\mathbf{L}}^{-1})_{c}%
^{a}\mathbf{d}(^{-}\widetilde{\mathbf{L}}\mathbf{)}_{b}^{c},\text{ }%
^{+}\widetilde{\mathbf{A}}_{b}^{a}=(^{+}\widetilde{\mathbf{L}}^{-1})_{c}%
^{a}\mathbf{d}(^{+}\widetilde{\mathbf{L}}\mathbf{)}_{b}^{c}\text{,}\nonumber
\end{align}
where%
\begin{align}
^{-}\widetilde{\mathbf{L}}_{_{b}}^{a}  &  =\delta_{b}^{a}-^{-}\omega_{b}%
^{a}\mathbf{m+\frac{\Lambda}{6}}\text{ }^{-}\mathbf{\Sigma}_{b}^{a}%
\mathbf{\mathbf{m}\overline{\mathbf{m}};}\text{ }^{+}\widetilde{\mathbf{L}%
}_{_{b}}^{a}=\delta_{b}^{a}-\text{ }^{+}\omega_{b}^{a}\overline{\mathbf{m}%
}-\mathbf{\frac{\Lambda}{6}}\text{ }^{+}\mathbf{\Sigma}_{b}^{a}%
\mathbf{\mathbf{m}\overline{\mathbf{m}};}\\
\widetilde{\mathbf{L}}_{b}^{a}  &  =\text{ }^{-}\widetilde{\mathbf{L}}_{c}%
^{a}\text{ }^{+}\widetilde{\mathbf{L}}_{b}^{c}=\text{ }^{+}%
\widetilde{\mathbf{L}}_{c}^{a}\text{ }^{-}\widetilde{\mathbf{L}}_{b}%
^{c}=\mathbf{L}_{b}^{a}+\mathbf{\frac{\Lambda}{6}(}^{-}\mathbf{\Sigma}_{b}%
^{a}-^{+}\mathbf{\Sigma}_{b}^{a})\mathbf{\mathbf{m}\overline{\mathbf{m}}%
.}\nonumber
\end{align}
and
\begin{equation}
\mathbf{L}_{b}^{a}=(\delta_{b}^{a}-^{-}\omega_{b}^{a}\mathbf{m)(}\delta
_{b}^{a}-\text{ }^{+}\omega_{b}^{a}\overline{\mathbf{m}}).
\end{equation}

Furthermore%
\begin{equation}
\theta^{a}=(\widetilde{\mathbf{L}}^{-1})_{b}^{a}\mathbf{d}%
\widetilde{\mathbf{x}}^{b}%
\end{equation}
where the same choice of generalized coordinates, $\widetilde{\mathbf{x}}%
^{a},$ can be made as in the vacuum case, that is $\widetilde{\mathbf{x}}%
^{a}=\mathbf{x}^{a}$ so that
\begin{equation}
\widetilde{\mathbf{x}}^{a}=-\frac{1}{2}\theta^{a}\mathbf{m-}\frac{1}{2}%
\theta^{a}\overline{\mathbf{m}}+\frac{1}{2}(^{+}\omega_{b}^{a}\theta
^{b}-\text{ }^{-}\omega_{b}^{a}\theta^{b})\mathbf{m}\overline{\mathbf{m}}.
\end{equation}

\section{Generalized Poincar\'{e} connections and Ricci flatness}

The results in the previous section can be re-expressed in terms of a
generalized Poincar\'{e} connection which is analogous to the ordinary
Poincar\'{e} connection, discussed in Section 2, with the ordinary forms in
Eq.(11) replaced by generalized forms.

A generalized Poincar\'{e} connection $_{p}\mathbf{A}$ is a generalized
one-form with values in the Lie algebra of the Poincar\'{e} group%
\begin{equation}
_{p}\mathbf{A}=\mathbf{e}^{a}V_{a}+\mathbf{a}_{b}^{a}V_{a}^{b}.
\end{equation}
with generalized curvature,%
\begin{equation}
_{p}\mathbf{F}=\mathbf{d}_{p}\mathbf{A}+\frac{1}{2}[_{p}\mathbf{A}%
,_{p}\mathbf{A}]=\mathbf{t}^{a}V_{a}+\text{ }\mathbf{f}_{b}^{a}V_{a}^{b}%
\end{equation}
where%
\begin{align}
\mathbf{t}^{a}  &  =\mathbf{de}^{a}+\text{ }\mathbf{a}_{b}^{a}\mathbf{e}%
^{b},\\
\mathbf{f}_{b}^{a}  &  =\mathbf{da}_{b}^{a}+\mathbf{a}_{c}^{a}\mathbf{a}%
_{b}^{c}.\nonumber
\end{align}
The differential forms $\mathbf{e}^{a}$, $\mathbf{t}^{a}$, $\mathbf{a}_{b}%
^{a}$, $\mathbf{f}_{b}^{a}$ are taken here to be, respectively, a generalized
co-frame, torsion two-form, connection one-form and curvature
two-form\footnote{Lower case Latin letters $\mathbf{e}^{a}$, $\mathbf{t}^{a}$,
$\mathbf{a}_{b}^{a}$, $\mathbf{f}_{b}^{a}$ are used here to describe any
Poincar\'{e} connection without specializing to the connections discussed
elsewhere.}.

The connection is flat when $_{p}\mathbf{F=0}$ and then%
\begin{align}
\mathbf{t}^{a}  &  =0\\
\mathbf{da}_{b}^{a}+\text{ }\mathbf{a}_{c}^{a}\text{ }\mathbf{a}_{b}^{c}  &
=0\nonumber
\end{align}
In a $5\times5$ matrix representation, $\mathbf{g}$, such a flat connection
can be expressed as $_{p}\mathbf{A}_{\beta}^{\alpha}\mathbf{=}$ ($\mathbf{g}%
^{-1})_{\gamma}^{\alpha}\mathbf{dg}_{\beta}^{\gamma}$, where
\begin{equation}
\mathbf{[g}_{\beta}^{\alpha}]=\left[
\begin{array}
[c]{cc}%
\mathbf{g}_{b}^{a} & \mathbf{g}^{a}\\
0 & 1
\end{array}
\right]
\end{equation}
($\alpha,\beta=1-5$), and $_{p}\mathbf{A}_{\beta}^{\alpha}$ has matrix
components corresponding respectively to $\mathbf{a}_{b}^{a}$ and
$\mathbf{e}^{a}$ given by%
\begin{equation}
_{p}\mathbf{A}_{b}^{a}=(\mathbf{g}^{-1})_{c}^{a}d\mathbf{g}_{b}^{c},\text{
}_{p}\mathbf{A}^{a}=(\mathbf{g}^{-1})_{b}^{a}\mathbf{dg}^{b}%
\end{equation}
where $\mathbf{g}_{b}^{a}$ and $\mathbf{g}^{a}$ are generalized zero-forms.
\ The matrix-valued generalized forms $\mathbf{g}$ constitute a (pointwise)
group, the generalized form analogue of the ordinary Poincar\'{e} group.

The matrix $\mathbf{g}$ is not unique since if $\widehat{\mathbf{g}%
}=\mathbf{kg}$ where the matrix-valued generalized form belonging to the
group, $\mathbf{k,}$is closed so that if
\begin{equation}
\lbrack\mathbf{k]=\mathbf{[k}_{\beta}^{\alpha}]=}\left[
\begin{array}
[c]{cc}%
\mathbf{k}_{b}^{a} & \mathbf{k}^{a}\\
0 & 1
\end{array}
\right]
\end{equation}
where $\mathbf{dk}_{b}^{a}=\mathbf{dk}^{a}=0$ then $\widehat{\mathbf{g}}%
^{-1}\mathbf{d}\widehat{\mathbf{g}}=\mathbf{g}^{-1}\mathbf{dg}$. \ In fact
$\widehat{\mathbf{g}}^{-1}\mathbf{d}\widehat{\mathbf{g}}=\mathbf{g}%
^{-1}\mathbf{dg}$ if and only if $\widehat{\mathbf{g}}=\mathbf{kg}$ where
$\mathbf{dk=0.}$

Consider now the particular generalized Poincar\'{e} connection where
$\mathbf{e}^{a}$ is an ordinary co-frame $\theta^{a}$, $\mathbf{a}_{b}^{a}%
$\textbf{ }is\textbf{ }$\mathbf{A}_{b}^{a}$ the generalized connection given
in Eq.(43) and where $\digamma_{b}^{a}$ is the curvature of the connection
$\alpha_{b}^{a}$, that is%
\begin{equation}
_{p}\mathbf{A}=\theta^{a}V_{a}+\mathbf{A}_{b}^{a}V_{a}^{b}=\theta^{a}%
V_{a}+(\alpha_{b}^{a}-\text{ }^{+}\digamma_{b}^{a}\overline{\mathbf{m}}-\text{
}^{-}\digamma_{b}^{a}\mathbf{m)}V_{a}^{b}.
\end{equation}
Since the generalized connection
\begin{equation}
\mathbf{A}_{b}^{a}=\alpha_{b}^{a}-^{-}\digamma_{b}^{a}\mathbf{m-}\text{ }%
^{+}F_{b}^{a}\overline{\mathbf{m}}%
\end{equation}
is flat by construction the generalized curvature of $_{p}\mathbf{A}$ is%
\begin{equation}
_{p}\mathbf{F}=[d\theta^{a}+\text{ }(\alpha_{b}^{a}-\text{ }^{-}\digamma
_{b}^{a}\mathbf{m-^{+}}\digamma_{b}^{a}\mathbf{\overline{\mathbf{m}})}%
\theta^{b}]V_{a}.
\end{equation}
\ Therefore $_{p}\mathbf{F=0}$ and the Poincar\'{e} connection $_{p}%
\mathbf{A}$ is flat if and only if $\alpha_{b}^{a}$ is the Levi-Civita
connection $\omega_{b}^{a}$ of the metric determined by the co-frame
$\theta^{a}$ and this metric is Ricci flat. \ It follows from the results of
the previous section, and those above, that this flat connection
$_{p}\mathbf{A}$ has a $5\times5$ matrix representation equal to
$\mathbf{g}^{-1}\mathbf{dg}$ where%

\begin{equation}
\mathbf{g}=[\mathbf{g}_{\beta}^{\alpha}]=\left[
\begin{array}
[c]{cc}%
\mathbf{L}_{b}^{a} & \mathbf{x}^{a}\\
0 & 1
\end{array}
\right]  =\left[
\begin{array}
[c]{cc}%
(\delta_{c}^{a}-^{+}\omega_{c}^{a}\overline{\mathbf{m}})(\delta_{b}^{c}%
-^{-}\omega_{b}^{c}\mathbf{m)} & \mathbf{x}^{a}\\
0 & 1
\end{array}
\right]
\end{equation}
and $($mod $\mathbf{du}^{a})$%
\begin{equation}
\mathbf{x}^{a}=-\frac{1}{2}\theta^{a}\mathbf{m-}\frac{1}{2}\theta^{a}%
\overline{\mathbf{m}}+\frac{1}{2}(^{+}\omega_{b}^{a}\theta^{b}-\text{ }%
^{-}\omega_{b}^{a}\theta^{b})\mathbf{m}\overline{\mathbf{m}}.
\end{equation}
In this matrix context the ordinary Poincar\'{e} gauge transformation given by
Eq.(14) results from the matrix multiplication $\mathbf{g\rightarrow g}_{1}=$
$\mathbf{g}\mathcal{P}$ where%
\begin{equation}
\mathcal{P}=\left[
\begin{array}
[c]{cc}%
L_{b}^{a} & \pi^{a}\\
0 & 1
\end{array}
\right]  .
\end{equation}
Under ordinary SO(1,3) gauge transformations, corresponding to setting
$\pi^{a}=0,$ $_{p}\mathbf{A}\rightarrow$ $_{p}\mathbf{A}_{1}=\theta_{1}%
^{a}V_{a}+(\omega_{1b}^{a}-$ $^{-}\Omega_{1b}^{a}\mathbf{m}-$ $^{+}\Omega
_{1b}^{a}\overline{\mathbf{m}}\mathbf{)}V_{a}^{b}$. \ Furthermore
$_{p}\mathbf{A}\mathbf{=g}^{-1}\mathbf{dg\rightarrow}_{P}\mathbf{A}_{1}=$
($\mathbf{g}_{1})^{-1}\mathbf{dg}_{1}$ where
\begin{align}
\mathbf{g}  &  \rightarrow\mathbf{g}_{1}=\mathbf{g}\mathcal{L}=\left[
\begin{array}
[c]{cc}%
\mathbf{L}_{b}^{a} & \mathbf{x}^{a}\\
0 & 1
\end{array}
\right]  \left[
\begin{array}
[c]{cc}%
L_{b}^{a} & 0\\
0 & 1
\end{array}
\right] \\
&  =\left[
\begin{array}
[c]{cc}%
(\delta_{c}^{a}-\text{ }^{+}\omega_{1c}^{a}\overline{\mathbf{m}})(\delta
_{b}^{c}-\text{ }^{-}\omega_{1b}^{c}\mathbf{m)} & \mathbf{x}_{1}^{a}\\
0 & 1
\end{array}
\right] \nonumber\\
\mathbf{x}_{1}^{a}  &  =-\frac{1}{2}\theta_{1}^{a}\mathbf{m-}\frac{1}{2}%
\theta_{1}^{a}\overline{\mathbf{m}}+\frac{1}{2}(^{+}\omega_{1b}^{a}\theta
_{1}^{b}-^{-}\omega_{1b}^{a}\theta_{1}^{b})\mathbf{m}\overline{\mathbf{m}%
}\nonumber
\end{align}
with $\theta_{1}^{a}=(L^{-1})_{b}^{a}\theta^{b},$ $^{+}\omega_{1c}%
^{a}\mathbf{=(}$ $^{+}L^{-1})_{b}^{a}d($ $^{+}L)_{c}^{b}+($ $^{+}L^{-1}%
)_{b}^{a}$ $^{+}\omega_{d}^{b}$ $^{+}L_{c}^{d}$ and analogously for
$^{-}\omega_{1c}^{a}$.

The$\ 5\times5$ matrix representations of $\mathbf{g,}$ given by Eqs.( 84 \&
85), can be factorized. \ It can be expressed as the product of two commuting
complex conjugate matrices, $\mathbf{h}$ and $\overline{\mathbf{h}},$ so that
\begin{equation}
\mathbf{g=h}\overline{\mathbf{h}}=\overline{\mathbf{h}}\mathbf{h.}%
\end{equation}
\ The matrix $\mathbf{h}$ is given by
\begin{equation}
\mathbf{h}=[\mathbf{h}_{\beta}^{\alpha}]=\left[
\begin{array}
[c]{cc}%
\mathbf{h}_{b}^{a} & \mathbf{h}^{a}\\
0 & 1
\end{array}
\right]  ,
\end{equation}
where%
\begin{align}
\mathbf{h}_{b}^{a} &  =^{-}\mathbf{L}_{b}^{a}=\delta_{b}^{a}-\text{ }%
^{-}\omega_{b}^{a}\mathbf{m}\\
\mathbf{h}^{a} &  =-\frac{1}{2}\theta^{a}\overline{\mathbf{m}}-\frac{1}%
{2}\text{ }^{-}\omega_{b}^{a}\theta^{b}\mathbf{m}\overline{\mathbf{m}%
}.\nonumber
\end{align}
Such matrices have values, pointwise, in a $5\times5$ matrix representation of
a sub-group of the complex Poincar\'{e} group (the semi-direct product of
$SL(2,\mathbb{C}$) and $\mathbb{C}^{4}$).

The flat generalized connection given by $_{P}^{-}\mathbf{A}=\mathbf{h}%
^{-1}\mathbf{dh}$ takes values in the Lie algebra of that group and has
non-zero components $\mathbf{a}_{b}^{a}=^{-}\mathbf{A}_{b}^{a}$ and
$\mathbf{e}^{a}=^{-}\mathbf{A}^{a}$ given by
\begin{align}
^{-}\mathbf{A}_{b}^{a}  &  =\text{ }^{-}\omega_{b}^{a}-^{-}\Omega_{b}%
^{a}\mathbf{m}\\
^{-}\mathbf{A}^{a}  &  =\frac{1}{2}\theta^{a}-\frac{1}{2}(\text{ }^{-}\tau
^{a})\overline{\mathbf{m}}-\frac{1}{2}[d(\text{ }^{-}\tau^{a})+\text{ }%
^{-}\omega_{b}^{a}\text{ }^{-}\tau^{b}]\mathbf{m}\overline{\mathbf{m}%
}\nonumber
\end{align}
where%
\begin{equation}
^{-}\tau^{a}=d\theta^{a}+\text{ }^{-}\omega_{b}^{a}\theta^{b}.
\end{equation}
Furthermore because the co-frame determines a solution of Einstein's vacuum
field equations and $\omega_{b}^{a}$ is the corresponding Levi-Civita
connection
\begin{align}
^{-}\tau^{a}  &  =-\text{ }^{+}\omega_{b}^{a}\theta^{b},\\
d(^{-}\tau^{a})+^{-}\omega_{b}^{a}\text{ }^{-}\tau^{b}  &  =\text{ }^{-}%
\Omega_{b}^{a}\theta^{b}=0\nonumber
\end{align}
and hence%
\begin{align}
^{-}\mathbf{A}_{b}^{a}  &  =\text{ }^{-}\omega_{b}^{a}-^{-}\Omega_{b}%
^{a}\mathbf{m}\\
^{-}\mathbf{A}^{a}  &  =\frac{1}{2}\theta^{a}+\frac{1}{2}(\text{ }^{+}%
\omega_{b}^{a}\theta^{b})\overline{\mathbf{m}}.\nonumber
\end{align}

Since the generalized connection $_{P}^{-}\mathbf{A}$ is flat it follows that
the pair\ $\mathbf{e}^{a}\equiv\equiv$ $^{-}\mathbf{A}^{a}$ and $\mathbf{a}%
_{b}^{a}\equiv$ $^{-}\mathbf{A}_{b}^{a}$ satisfy the generalized Cartan
structure equations, Eq.(76), with $\mathbf{t}^{a}$= $\mathbf{f}_{b}^{a}=0$,
in particular%
\begin{equation}
\mathbf{de}^{a}+(\text{ }^{-}\omega_{b}^{a}-^{-}\Omega_{b}^{a}\mathbf{m)e}%
^{b}=0.
\end{equation}
. \ Similar results follow in the obvious way for the complex conjugates
$\overline{\mathbf{h}}$, $_{P}^{+}\mathbf{A}=\overline{\mathbf{h}}%
^{-1}\mathbf{d}\overline{\mathbf{h}}$ and their components.

Since $\mathbf{g=}\overline{\mathbf{h}}\mathbf{h}$ it follows that the
relationship between these flat connections is given by%
\begin{equation}
_{P}\mathbf{A=}_{P}^{-}\mathbf{A+h}^{-1}\text{ }_{P}^{+}\mathbf{Ah}%
\end{equation}

Because any two solutions of Einstein's vacuum field equations, with
respective co-frames $\theta_{1}^{a}$ and $\theta_{2}^{a}$ and Levi-Civita
one-forms $\omega_{1b}^{a}$ and $\omega_{2b}^{a},$ can be associated with
corresponding matrices $\mathbf{g}_{1}$ and $\mathbf{g}_{2}$, as in Eqs.(84 \&
85), $\mathbf{g}_{2}$ will be related to $\mathbf{g}_{1}$ by a matrix-valued
generalized form $\mathbf{c}$. \ Right or left multiplication can be chosen
with the former illustrated here. \ Choosing right multiplication, so that
$\mathbf{g}_{2}=$ $\mathbf{g}_{1}\mathbf{c}$, the corresponding generalized
connection one-forms $_{p}\mathbf{A}_{2}$ and $_{p}\mathbf{A}_{1}$ are then
related by a generalization of a Poincar\'{e} gauge transformation%
\begin{equation}
_{p}\mathbf{A}_{2}=\mathbf{c}^{-1}\text{ }_{p}\mathbf{A}_{1}\mathbf{c+c^{-1}%
\mathbf{dc,}}%
\end{equation}
where%
\begin{equation}
\mathbf{c=}\left[
\begin{array}
[c]{cc}%
\mathbf{c}_{b}^{a} & \mathbf{c}^{a}\\
0 & 1
\end{array}
\right]
\end{equation}
and%
\begin{align}
\mathbf{c}_{b}^{a}  &  =(\delta_{c}^{a}-^{+}\mu_{c}^{a}\overline{\mathbf{m}%
})(\delta_{b}^{c}-^{-}\mu_{b}^{c}\mathbf{m)}\\
^{+}\mu_{c}^{a}  &  =^{+}\omega_{2c}^{a}-\text{ }^{+}\omega_{1c}^{a};\text{
}^{-}\mu_{b}^{c}=^{-}\omega_{2b}^{c}-\text{ }^{-}\omega_{1b}^{c};\text{
}\nonumber\\
\mathbf{c}^{a}  &  =(\mathbf{L}_{1b}^{a})^{-1}(\mathbf{x}_{2}^{b}%
-\mathbf{x}_{1}^{b}).\nonumber
\end{align}

In concluding this section it is briefly noted that when Einstein's field
equations with a zero energy-momentum tensor but non-zero cosmological
constant are considered analogous results hold. \ There is again a flat
generalized Poincar\'{e} connection given, using the notation of Section 4, by%
\begin{equation}
_{p}\widetilde{\mathbf{A}}=\theta^{a}V_{a}+\widetilde{\mathbf{A}}_{b}^{a}%
V_{a}^{b}.
\end{equation}
Following the same procedure as in the vacuum case a five dimensional
representation of the flat connection $_{p}\widetilde{\mathbf{A}}$ is given by
$\widetilde{\mathbf{g}}^{-1}\mathbf{d}\widetilde{\mathbf{g}}$ where%
\begin{equation}
\widetilde{\mathbf{g}}_{\beta}^{\alpha}=\left[
\begin{array}
[c]{cc}%
\widetilde{\mathbf{L}}_{b}^{a} & \widetilde{\mathbf{x}}^{b}\\
0 & 1
\end{array}
\right]  ,
\end{equation}
and $\widetilde{\mathbf{L}}_{b}^{a}$ and $\widetilde{\mathbf{x}}^{b}$ are
given by Eqs.(70 \& 73).

As in the vacuum case, but now in this broader context where $\Lambda$ is not
necessarily zero but $T_{ab}=0$, it follows that any two solutions of
Einstein's equations, determined respectively by co-frames $\theta_{1}^{a}$
and $\theta_{2}^{a}$ and Levi-Civita one-forms $\omega_{1b}^{a}$ and
$\omega_{2b}^{a},$ will have corresponding matrices $\widetilde{\mathbf{g}%
}_{1}$ and $\widetilde{\mathbf{g}}_{2}$, as in Eq.(101), related by a
matrix-valued generalized form $\widetilde{\mathbf{c}}$. \ Right or left
multiplication can be chosen. \ Choosing right multiplication as before, so
that $\widetilde{\mathbf{g}}_{2}=$ $\widetilde{\mathbf{g}}_{1}%
\widetilde{\mathbf{c}}$, the corresponding generalized connection one-forms
$_{p}\widetilde{\mathbf{A}}_{2}$ and $_{p}\widetilde{\mathbf{A}}_{1}$ are
related by a generalization of a Poincar\'{e} gauge transformation.

\section{An action principle using a Nieh-Yan form}

Generalized Chern-Simons action principles for Einstein's field equations were
constructed in \cite{rob9} by using type $N=2$ generalized connections which
were flat when the Euler-Lagrange equations were satisfied,. \ However these
action principles did not include the vacuum case. \ In this section a
connection with these properties is used in a generalization of \ the Nieh-Yan
three-form.\ \ This three-form, rather than a generalized Chern-Simons
three-form, is then used as a first order action, with co-frames and
connection one-forms as independent variables, for Einstein's vacuum field equations.

The ordinary Nieh-Yan three-form is
\begin{equation}
NY=\theta_{a}\mathsf{D}\theta^{a}=\theta_{a}\Theta^{a}%
\end{equation}
where, as in Section 2, $\theta^{a}$ is an orthonormal co-frame, $D$ denotes
the exterior covariant derivative with respect to an ordinary connection
$\alpha_{b}^{a}$ with values in the Lie algebra of the Lorentz group and
$\Theta^{a}$ is the ordinary torsion two-form.

The exterior derivative of $NY$ is the well-known Nieh-Yan four-form $N$
\cite{N-Y}, \cite{Li} where%
\begin{equation}
N=d(\theta_{a}D\theta^{a})=\Theta_{a}\Theta^{a}-\theta_{a}\theta^{b}F_{b}^{a},
\end{equation}
and $F_{b}^{a}$ is the curvature of $\alpha_{b}^{a}$.

The generalized Nieh-Yan three-form, $\mathbf{NY}$, is constructed by
replacing the ordinary connection form $\alpha_{b}^{a}$ by the generalized
connection form $\mathbf{A}_{b}^{a}$ introduced in Eq.(40)%
\begin{equation}
\mathbf{A}_{b}^{a}=\alpha_{b}^{a}-\digamma_{b}^{a}\mathbf{m}^{1}+\text{
}^{\ast}\digamma_{b}^{a}\mathbf{m}^{2},
\end{equation}
where $\digamma_{b}^{a}$ is the curvature of $\alpha_{b}^{a}$, so that the
generalized connection is flat. Then\
\begin{equation}
\mathbf{NY}=\theta_{a}\mathbf{D}\theta^{a}.
\end{equation}
and the generalized Nieh-Yan four-form, $\mathbf{N}$, is%
\begin{equation}
\mathbf{N}=\mathbf{d}(\theta_{a}\mathbf{D}\theta^{a}).
\end{equation}
Now let $\mathbf{c}_{3}$ be a 3-polychain as in Section 3 and consider the
action integral $\mathbf{I}$ constructed by integrating the generalized
Nieh-Yan three-form over $\mathbf{c}_{3}$%
\begin{equation}
\mathbf{I}=\int_{\mathbf{c}_{3}}\theta_{a}\mathbf{D}\theta^{a}.
\end{equation}%
\begin{equation}
\mathbf{I=}\int_{c_{3}}\theta_{a}D\theta^{a}+\int_{_{c_{4}}^{1}}\digamma
_{b}^{a}\theta_{a}\theta^{b}-\int_{_{c_{4}}^{2}}\text{ }^{\ast}\digamma
_{b}^{a}\theta_{a}\theta^{b}%
\end{equation}

After expanding in terms of ordinary forms and chains the variation of
$\mathbf{I}$ is
\begin{align}
\delta\mathbf{I}  &  \mathbf{=\{-}\int_{\partial c_{3}}\theta_{a}\delta
\theta^{a}+\int_{c_{3}}(2\delta\theta_{a}D\theta^{a}-\delta\alpha_{ab}%
\Sigma^{ab})\\
&  +\int_{\partial_{c_{4}}^{1}}\delta\alpha_{ab}\Sigma^{ab}+\int_{_{c_{4}}%
^{1}}(2\delta\theta_{a}F_{b}^{a}\theta^{b}+\delta\alpha_{ab}D\Sigma^{ab})\\
&  -\int_{\partial_{c_{4}}^{2}}[(\delta\alpha_{ab}(^{\ast}\Sigma^{ab}%
)]-\int_{_{c_{4}}^{2}}[2\delta\theta_{a}\text{ }^{\ast}\digamma_{b}^{a}%
\theta^{b}+\delta\alpha_{ab}D(^{\ast}\Sigma^{ab})]\}.\nonumber
\end{align}
where $\Sigma^{ab}=\theta^{a}\theta^{b}$, $^{\ast}\Sigma^{ab}=\frac{1}%
{2}\epsilon_{abcd}\theta^{c}\theta^{d}$ and $^{\ast}\digamma_{b}^{a}=\frac
{1}{2}\epsilon_{bcd}^{a}\digamma^{cd}$.

A specific interpretation of the vanishing of $\delta\mathbf{I}$ depends on
the choice of polychain and the boundaries\footnote{It is assumed that the
relevant variables and their pull-backs are defined on the chains and their
boundaries.}. \ However the vanishing of the integral over $_{c_{4}}^{2}$ for
arbitrary independent variations $\delta\alpha_{ab}$ and $\delta\theta_{a}$
implies that $D(^{\ast}\Sigma^{ab})=0$ and $^{\ast}\digamma_{b}^{a}\theta
^{b}=0$ there. \ In other words $\alpha_{b}^{a}$ is the Levi-Civita spin
connection determined by the co-frame $\theta^{a}$ and the metric $ds^{2}%
=\eta_{ab}\theta^{a}\otimes\theta^{b}$ satisfies Einstein's vacuum equations there.

\section{Discussion and Conclusions}

A novel approach to Einstein's gravitational field equations for Lorentzian
signature metrics in four space-time dimensions has been developed. \ Type
$N=2$ generalized forms were employed and Einstein's equations were
represented by extensions of Cartan's structure equations for metric
geometries. \ When the energy-momentum tensor vanished co-frames for all such
solutions were shown to be describable in a "flat" form, that is by four
generalized one-forms $\theta^{a}=(\mathbf{L}^{-1})_{b}^{a}\mathbf{dx}^{b}$.
\ It was demonstrated that these solutions determine, and are determined by,
flat generalized connections with values the Lie algebra of the Poincar\'{e}
group. \ Hence all solutions are related by generalized gauge transformations.
\ A factorization which was used to relate complex conjugate flat connections
to these flat generalized connections was explicitly exhibited in the vacuum
case. The encoding of Einstein's field equations in a generalization of
Cartan's equations was extended to the case where the matter energy-momentum
tensor, $T_{ab}$, was non-zero. \ It was shown that the energy-momentum tensor
(and cosmological constant $\Lambda)$ could either be represented by a
generalized torsion two-form and a flat generalized connection or could be
added to the metric connection used in the vacuum case to constitute a new
metric connection. \ 

Past use of type $N=2$ generalized forms to construct action principles for
the $\Lambda\neq0$ Einstein equations and Yang-Mills fields \cite{rob9} was
extended. \ In previous work the action principles were constructed by using
generalized Chern-Simons three-forms. Here an action principle for the vacuum
Einstein equations was constructed by making use of a generalization of the
Nieh-Yan three-form in which the ordinary Levi-Civita connection is replaced
by a flat generalized connection.

The general approach, and new perspectives, developed in this work can be
applied to four-metrics with other signatures and to holomorphic four-metrics
and half-flat metrics. \ They could also be used to study systems with
non-zero energy-momentum tensors such as the Einstein-Maxwell system of
equations. \ The source-free Yang-Mills equations can be represented by flat
type $N=2$ generalized connections, see e.g.\cite{rob4}, and can also be
explored along the lines pursued in this paper.

Aspects of some of the results presented in this paper can be related to
certain research on higher gauge theories, \cite{song1}, \cite{song2}, and
recall, although it is not the same, work on teleparallelism, see for example,
,\cite{tel}, \cite{baez}.

\textbf{Acknowledgement: \ }I would like to thank Chrys Soteriou for reading
an earlier version of this paper and for drawing my attention to the research
on teleparallelism.

\end{document}